\documentclass[numbered]{sn-jnl}
\usepackage[utf8]{inputenc}
\usepackage{graphicx}
\usepackage{multirow}
\usepackage{amsmath,amssymb,amsfonts,mathtools}
\usepackage{mathrsfs}
\usepackage{xcolor}
\usepackage{textcomp}
\usepackage{physics}
\usepackage{manyfoot}
\usepackage{natbib}
\usepackage{algorithm}
\usepackage{algorithmicx}
\usepackage{algpseudocode}
\usepackage{subcaption}
\usepackage{listings}

\graphicspath{{Figures/}}
\begin{document}

\title[Multi-domain analysis and prediction of the light emitted by an inductively coupled plasma jet]{Multi-domain analysis and prediction of the light emitted by an inductively coupled plasma jet}

\author*[1,2]{\fnm{Lorenzo} \sur{Capponi}}\email{lcapponi@illinois.edu}
\author[1,2]{\fnm{Alberto} \sur{Padovan}}\email{padovan3@illinois.edu}
\author[1,2]{\fnm{Gregory S.} \sur{Elliott}}\email{elliottg@illinois.edu}
\author[1,2]{\fnm{Marco} \sur{Panesi}}\email{mpanesi@illinois.edu}
\author[1,2]{\fnm{Daniel J.} \sur{Bodony}}\email{bodony@illinois.edu}
\author[1,2]{\fnm{Francesco} \sur{Panerai}}\email{fpanerai@illinois.edu}

\affil[1]{\orgdiv{Department of Aerospace Engineering}, \orgname{University of Illinois at Urbana-Champaign}, \orgaddress{\street{104 S. Wright St.}, \city{Urbana}, \postcode{61801} \state{IL}}}

\affil[2]{\orgdiv{Center for Hypersonics and Entry Systems Studies}, \orgname{University of Illinois at Urbana-Champaign}, \orgaddress{\street{105 S. Goodwin Ave.}, \city{Urbana}, \postcode{61801} \state{IL}}}

\abstract{
Inductively coupled plasma wind tunnels are crucial for replicating hypersonic flight conditions in ground testing. 
Achieving the desired conditions (e.g., stagnation-point heat fluxes and enthalpies during atmospheric reentry) requires a careful selection of operating inputs, such as mass flow, gas composition, nozzle geometry, torch power, chamber pressure, and probing location along the plasma jet. 
The study presented herein focuses on the influence of the torch power and chamber pressure on the plasma jet dynamics within the 350 kW Plasmatron X ICP facility at the University of Illinois at Urbana-Champaign.
A multi-domain analysis of the jet behavior under selected power-pressure conditions is presented in terms of emitted light measurements collected using high-speed imaging. 
We then use Gaussian Process Regression to develop a data-informed learning framework for predicting Plasmatron X jet profiles at unseen pressure and power test conditions.
Understanding the physics behind the dynamics of high-enthalpy flows, particularly plasma jets, is the key to properly design material testing, perform diagnostics, and develop accurate simulation models. 
}

\maketitle

\section{Introduction}\label{introduction}
Hypersonic space vehicles operate under extreme flight conditions, subjecting their thermal protection systems (TPS) to complex and demanding challenges~\citep{duffa2013ablative, anderson2000hypersonic}. The integrity and performance of TPS play a crucial role in safeguarding the vehicle and its payload from the harsh aerothermal, chemical, and mechanical processes experienced during flight~\citep{gnoffo1999planetary, park1989nonequilibrium}. For example, within the shock layer of hypersonic systems, the gas flow is dominated by non-equilibrium thermochemistry and radiation, leading to the generation of a plasma layer in the post-shock region through gas dissociation and ionization~\citep{anderson2000hypersonic}. Consequently, the interaction between the plasma and the TPS surface governs the material response and the performance during hypersonic entries~\citep{gilmore2002spacecraft}. 
To investigate and describe these effects, the community has developed experimental platforms capable of capturing and replicating these phenomena~\citep{calomino2010evaluation, balter1992arc, purpura2008experimental, loehle2022assessment}. 
Among these facilities, inductively coupled plasma (ICP) wind tunnels are used to reproduce the desired hypersonic flight conditions in a near-continuous, controlled and chemically-pristine environment, achieving targeted stagnation-point cold-wall heat fluxes and enthalpies~\citep{fletcher2006characterization, bottin1999vki,capponi2023aerothermal}. 

This research focuses on the 350 kW Plasmatron X ICP wind tunnel, commissioned in early 2022 by the Center for Hypersonics and Entry Systems Studies (CHESS), at the University of Illinois at Urbana-Champaign~\citep{capponi2023aerothermal, oldham2023aerothermal}. 
At the Plasmatron X, the desired entry conditions can be accomplished through an appropriate selection of operating input settings and parameters, such as mass flow and gas composition, ICP torch power, reactor chamber static pressure, nozzle geometry (e.g., straight, contoured or conical converging-diverging), and probing/testing location along the plasma jet~\citep{lieberman1994principles, fridman2008plasma}. For a given mass flow rate, the power supplied to the ICP torch and the reactor static pressure are two dominant factors in tailoring the behavior of the plasma jet~\citep{capponi2023aerothermal, oldham2023aerothermal, lieberman1994principles}. 
Effectively, torch power and chamber pressure determine the aerothermochemical properties and jet dynamics, altering plasma density and viscosity, emitted radiation, and flow field characteristics~\citep{clemens1995large,papamoschou1988compressible}. 
The combination of these properties lead to the generation of supersonic conditions, jet instabilities, and distinctive features (e.g., shock location, plasma jet length and core width) whose detailed knowledge is imperative to effectively design material test campaigns, perform experimental diagnostics, and inform numerical simulation models~\citep{panesi2007analysis, kumar2022self,oruganti2023modeling, jo2023multi, jo2023validation, abeele2000efficient, degrez2004numerical,zhang2016analysis}.

For the purpose of studying flow characteristics, particularly in high-temperature flows, image-based techniques and non-intrusive measurements are usually preferred over traditional intrusive approaches, as these can disrupt the flow, leading to inaccuracies and incomplete data~\citep{raghu1995visualization, kurelek2023superposition}. 
Advanced image-based methods offer high-resolution visualization of complex flow phenomena, providing insight into turbulence and energy transfer in extreme environments without perturbing the delicate nature of the system~\citep{vitkovicova2020identification,geschwindner2022ultra, oka2023experimental, cakir2023assessment}.

Different approaches have been presented to study the free stream plasma jet produced by ICP wind tunnels. \cite{cipullo2014investigation} used high-speed imaging to associate jet unsteadiness spatial distribution, in the frequency domain, to the electrical features of the radio-frequency system. 
\cite{zander2017high} employed high-speed imaging techniques to compare the high-frequency effects of a CO$_2$ plasma flow against laser-diagnostics-based data from previous studies~\citep{marynowski2014aerothermodynamic}. 
Similarly, \cite{fagnani2020investigation} used high-resolution optical emission spectroscopy together with magneto-hydrodynamics simulations to investigate the properties of an ICP plasma jet. Finally, \cite{fries2022time} proposed a high-speed imaging and phased-averaged optical emission spectroscopy approach to investigate time-dependent variations in plasma temperature in both the exit plasma plume and the plasma core, for argon and air plasmas. 
As far as numerical studies are concerned, \cite{anfuso2021multiscale} and \cite{demange2018absolute, demange2020local, demange2020role, demange2022global} investigated ICP jets, with a focus on instabilities and coherent structures.
Despite the large body of experimental and numerical work on ICP jets, to the best of the authors' knowledge, a thorough experimental study of the jet dynamics as a function of operating conditions (e.g., torch power and chamber pressure) is missing in the literature. It is the focus of the work presented herein to start bridging this gap. 

The contribution of this work is two-fold. First, the Plasmatron X jet behavior and dynamics are investigated in order to describe features such as subsonic-to-supersonic transitions and shock diamond presence and location, and to identify fluid dynamics effects (e.g., unsteadiness and instabilities) as a function of delivered torch power and reactor chamber pressure. 
This is achieved through high-speed image-based measurements of the light emitted from the Plasmatron X plasma jet, where the jet centerline profiles are processed and analyzed in the space, time, and frequency domains.  
Second, we demonstrate that the well-known machine learning paradigm of Gaussian Process Regression (GPR)~\citep{williams1995,williams2006gaussian} can be used to fit the jet centerline profiles as a function of pressure and power. 
More specifically, given the low-rank nature of the data, we first use Proper Orthogonal Decomposition (POD) (also known as Principal Component Analysis)~\citep{lumley1967,sirovich1987turbulence,berkooz1993} to identify the spatial modes of largest variance, and then we apply GPR to fit the POD coefficients.
This procedure delivers a data-driven model that is capable of estimating the jet profiles at unseen pressure-power combinations, and this is particularly useful especially given the cost and extensive data-acquisition times associated with running the ICP facility in its extensive operational envelope. 

The manuscript is organized as follows. 
In Sec.~\ref{sec:methods} we first describe the experimental setup, the data processing, and methodologies used to study the jet behavior and its dynamics, and we also describe our use of POD and GPR to predict unseen conditions.
In Sec.~\ref{sec:results}, we then present and discuss the results. 
Sec.~\ref{sec:conclusions} draws the conclusions.

\section{Methodology}\label{sec:methods}
\subsection{Plasmatron X facility}
\begin{figure}
        \centering        
        \includegraphics[width=1\textwidth]{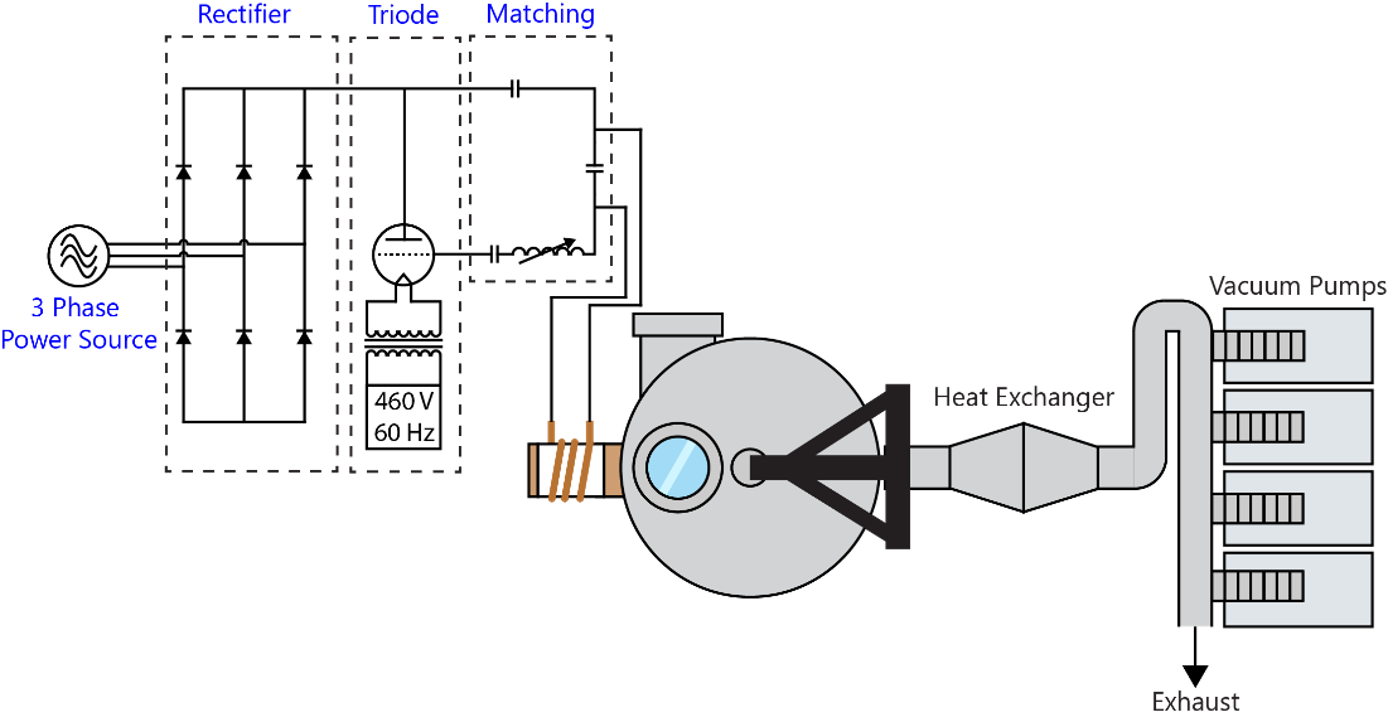}
        \caption{Schematic of the Plasmatron X facility.}
        \label{fig:PX}
\end{figure}

The ICP torch wind tunnel at the Plasmatron X facility was designed and engineered by CHESS and Tekna Plasma Systems, Inc. (Sherbrooke, QC, Canada). A simplified schematic of the Plasmatron X is shown in Fig.~\ref{fig:PX}.
A 12-pulse rectifier operating in three phases is utilized to transform the standard industrial line voltage (460 V${rms}$, 60 Hz) into a high voltage output of up to 15 kV${dc}$. This high voltage is then directed through a vacuum triode to generate radiofrequency (RF) power.
To ensure proper impedance matching, variable capacitors and inductors are adjusted based on the specific nozzle configuration and operational conditions. The plasma generation process involves applying an RF excitation frequency of 2.1 MHz to a 3-turn induction coil wrapped around a ceramic tube with an inner diameter of 100 mm. This setup is capable of operating at input power levels ranging from 13.5 to 350 kW.
Gas is introduced into the torch body through both central and sheath gas lines, with the sheath gas being injected at a 15-degree angle to induce vorticity in the flow. The RF current applied to the coil generates an induced magnetic field, which is absorbed by the electrons and ions within the plasma through electromagnetic induction. The increased collision rate resulting from the swirling gas feed contributes to Joule heating, resulting in the production of a stable, high-temperature plasma jet.

Following its exit from the torch, the plasma jet enters a cylindrical stainless-steel vacuum chamber with an inner diameter of 1.2 m and a length of 1.8 m. The chamber maintains a base pressure of 40 Pa and can be controlled to reach pressures of up to 150 kPa through a valve manifold and four high-capacity dry screw pumps. The high-temperature flow is then directed into a heat exchanger via a catch-cone diffuser, and the exhaust gases are released into the atmosphere through a dedicated purging system. This allows for adjustments to the flow properties during operation to accurately replicate changes that occur during hypersonic flight profiles. These adjustments can be made by modifying various system settings such as ICP power, chamber pressure, mass flow rate, gas composition, and nozzle geometry.
More details on the ICP wind tunnel performances and configurations are given in~\cite{capponi2023aerothermal} and~\cite{oldham2023aerothermal}. 

\subsection{Plasma jet behavior}\label{sec:jet_profile}
\begin{figure}
        \centering        
        \includegraphics[width=0.7\textwidth]{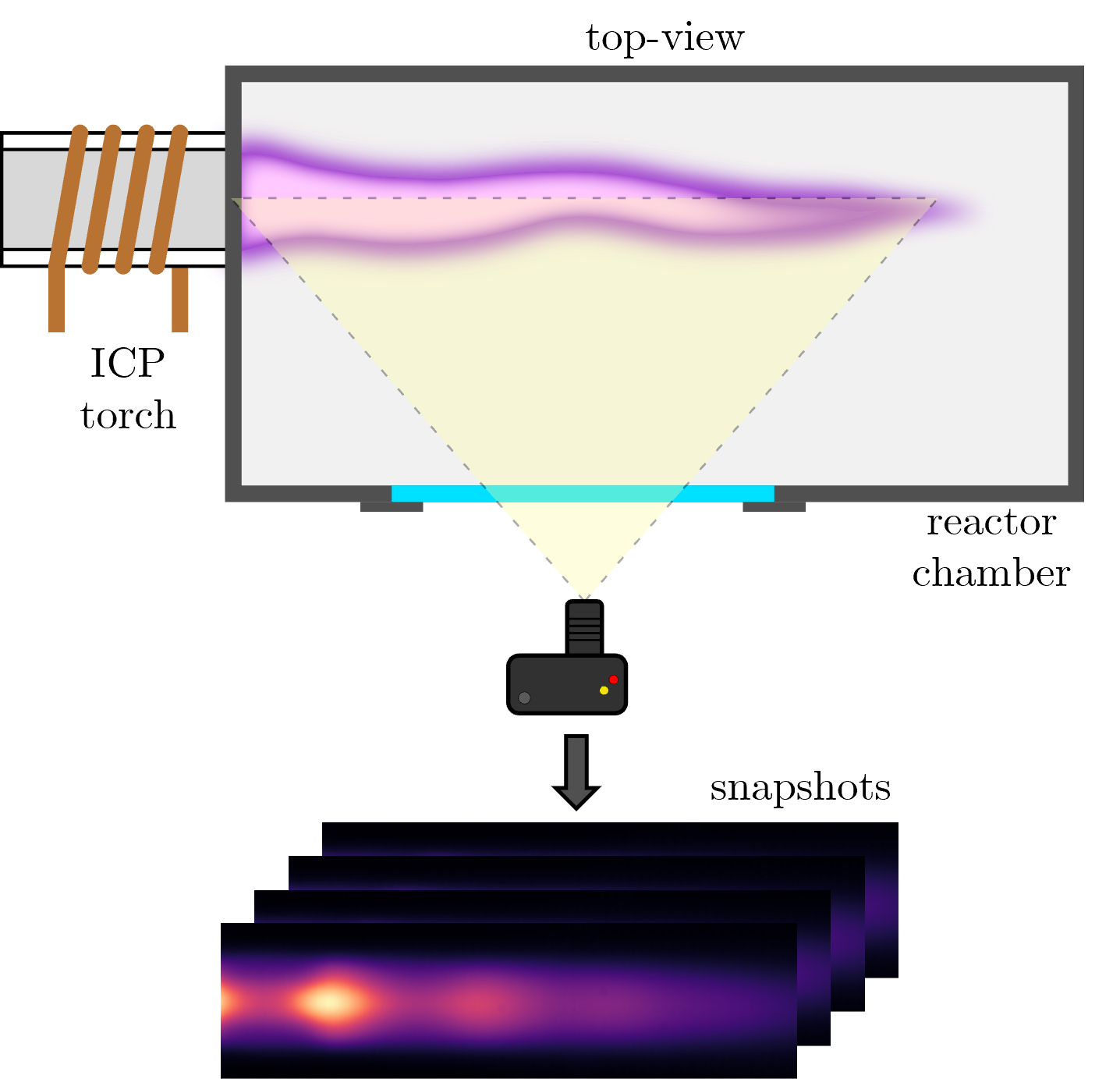}
        \caption{Schematic of experimental setup at Plasmatron X wind tunnel: high-speed camera acquiring the plasma jet emitted light through a quartz viewport.}
        \label{fig:setup}
\end{figure}

\begin{figure}
        \centering        
        \includegraphics[width=0.8\textwidth]{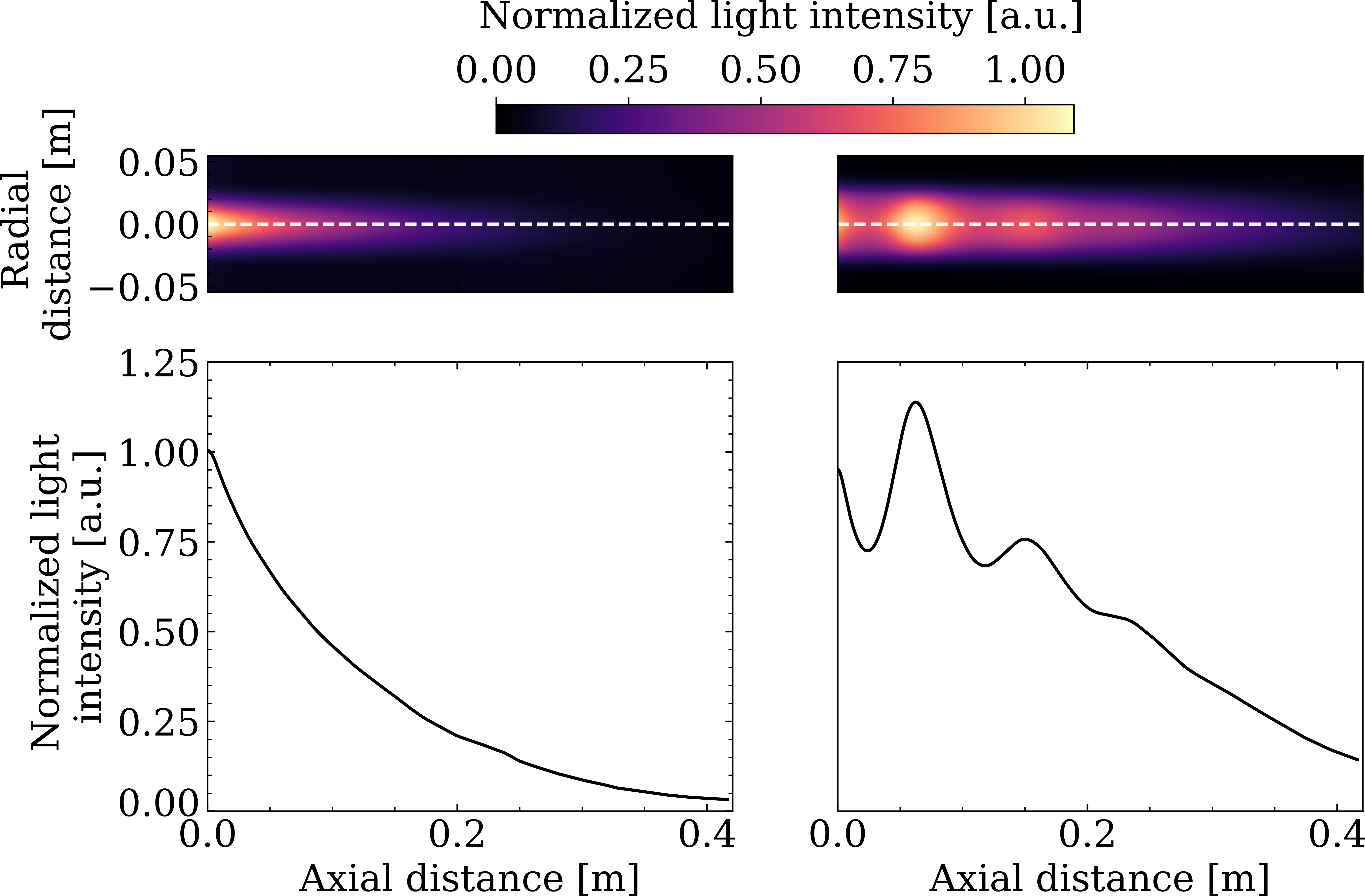}
        \caption{Normalized and filtered light intensity jet centerline (bottom), extracted from a snapshot (top) at 600 Pa and 60 kW (left) and 200 kW (right) test cases.}
        \label{fig:maps_and_prof}
\end{figure}

At a given chamber pressure $P$ and torch power $W$, measurements of the emitted light field of the jet discharge are taken using the monochromatic sensor of a Kron Chronos 2.1-HD camera (12-bit resolution, 350-900 nm spectral range), mounting a Micro Nikkor 60 mm f/2.8D lens (see schematic in Fig.~\ref{fig:setup}).
The equivalent radiometric quantity that is measured is the irradiance $E_e$, radially-integrated over the line of sight and over the wavelength range allowed by the camera sensor~\citep{griffiths2005introduction}.
In fluid dynamics, irradiance reveals flow properties, enabling flow visualization, property change detection, and heat transfer measurement. Monitoring irradiance is essential for understanding flow aspects like velocity, turbulence, particle concentration, and stability.
However, since the computation of $E_e$ in units of [W/m$^2$] requires calibrating the collected data against a photometric light source, we instead present the measurements in terms of emitted light intensity with units [counts/m$^2$], where the area to be considered is the pixel size area. 
Therefore, throughout our analysis, we consider the plasma jet emitted light scalar-valued field $L(x, y, t, P, W)$, where $x$ and $y$ are the axial and radial coordinates, and $t$ is the acquisition time.
For each experimental measurement, the sampling frequency of the camera is set to 1 kHz at a resolution of 1850~$\times$~480 pixels, and data are collected over 1 second. 
When higher temporal resolution is required to resolve the jet dynamics, a second acquisition set is taken at 10 kHz over a shorter time interval of 0.1 s, by reducing the spatial resolution along the radial dimension. 
As the exposure time and aperture are adapted to each particular test case to accommodate for the variable brightness of the plasma jet, an overall intensity normalization is performed as a pre-processing step, while also ensuring that relative emitted light information (across acquisitions) would not be lost. The normalization process relies on a reference area located at the torch exit (i.e., axial location $x$ = 0) within the reactor chamber. 
Ideally, the light intensity captured by the camera at this reference point should remain consistent across various facility and acquisition settings. Therefore, we assign a value of 1 to the reference point, serving as the anchor for the overall intensity normalization.
A spatial calibration of the camera setup is applied using a calibration target, and the pixel size is determined to be approximately 0.217 mm for the present campaign. Preliminary studies revealed that most of the information lives on the jet centerline~\citep{oldham2023aerothermal}.
For this reason, the axial profile at the centerline is extracted from two-dimensional measurements, simplifying the emitted light field function to $L(x, t, P, W)$. Fig.~\ref{fig:maps_and_prof} shows normalized light intensity distribution maps, with jet centerline profiles (dashed lines), for a subsonic and supersonic jet conditions.
Normalized centerline profiles are also processed by applying a Gaussian spatial filter in order to remove acquisition noise. 
A series of pressure and power conditions is selected based on previous studies~\citep{capponi2023aerothermal}, and steady-state emitted light snapshots are collected and pre-processed as previously described, for a total of 73 experimental combinations, with $P\in [300,10000]$ Pa, and $W\in [50,300]$ kW. All the experiments are performed using a straight nozzle (ID = 100 mm), with a constant air mass flow of 8 g/s~\citep{capponi2023aerothermal}.
The time-averaged centerline profiles $\Bar{L}(x, P, W)$ are used to describe the emitted light intensity distribution as a function of torch power and pressure chamber for the entire jet axial length, and for predicting steady-state jet profiles for unseen power-pressure conditions, as presented in Sec.~\ref{sec:gpr_method}.

Higher-speed acquisitions are used to evaluate fluctuations of emitted light to identify fluid dynamics effects and instabilities due to pressure-power conditions. 
For fixed pressures $P$ and powers $W$, two space-time data processing methods are used. A standard two-point space-time correlation analysis is used in its Eulerian form as~\citep{he2017space}
\begin{equation}
    R_{LL} (x,t,\xi,\tau) = \frac{\left \langle L(x,t)\,L(x+\xi, t + \tau)\right \rangle}{\sqrt{\left \langle L^2(x,t)\right \rangle\left \langle L^2(x+\xi, t + \tau)\right \rangle}},
    \label{eq:stcorr}
\end{equation}
where $\xi$ and $\tau$ are spatial and temporal lags, respectively, and $\langle\cdot\rangle$ denotes the expectation operator.
Similarly, the emitted light fluctuation $L^\prime(x,t)$ across the time-averaged profile is defined as
\begin{equation}
    L^\prime(x,t) =  L(x,t) - \Bar{L}(x).
    \label{eq:prime}
\end{equation}
Separately, a frequency domain-based approach can be used to quantify the dominant spectral components of the dynamics of the plasma jet. The Short-Time-Fourier-Transform (STFT) at specific axial locations is used to examine the time-varying frequency components, and identifying specific features and transients of the emitted light intensity~\citep{shin2008fundamentals}. In the discrete-time case, the data are divided into windows (with a non-zero overlap in order to reduce spectral leakage) and then they are Fourier-transformed. The complex result is then added to a matrix, which records magnitude and phase for each point in time and frequency. 
For a given pressure $P$ and power $W$, at axial location $x = 0$, the discrete STFT of $L^\prime(t)$ can be expressed by~\citep{sejdic2009time}
\begin{equation}
\textrm{STFT}\{L^\prime \}(q, \omega) = \sum_{t=0}^{T-1} L^\prime\left(t\right)  w\left(t - q\right) e^{-\textrm{i} \omega t},
\end{equation}
with a discrete window-function $w$, and total number of samples $T$ in the signal~$L^\prime$~\citep{virtanen2020scipy}.

\subsection{Prediction of emitted light profiles for unseen operating conditions}\label{sec:gpr_method}
In this section, we illustrate how GPR and POD are used to predict time-averaged jet centerline emitted light profiles at unseen pressures and powers. 
Henceforth, let $\ell(t,C)$ be an $M$-vector that denotes the spatially-discrete jet centerline emitted light profile, where $M$ is the number of pixels (spatial locations) in the axial direction, and $C \coloneqq (P,W)$ denotes the independent parameters of the problem. 
The time-averaged emitted light profile is defined as
\begin{equation}
    \overline{\ell}(C) = \frac{1}{T}\int_0^T \ell(t,C)\,dt,
\end{equation}
where $T$ is the length of the data-acquisition interval. 
Given $\overline{\ell}$ over a training set $\{C_i\}_{i=1}^N$, we want to learn the functional dependence of $\overline{\ell}$ on the parameters $C$ for predicting the time-averaged jet emitted light at unseen pressure-power combinations. 

Since $M$ is large (i.e., 1850 pixels in the centerline), we first rely on POD for a reduced-order representation of the time-averaged emitted light data.
In particular, given the data matrix $\mathbf{X} = \{\overline{\ell}(C_i)\} \in \mathbb{R}^{M\times N}$, we seek an orthonormal matrix $\mathbf{U}\in\mathbb{R}^{M\times r}$ that minimizes the objective
\begin{equation}
\label{eq:pod_obj}
    \lVert \left(\mathbf{I} - \mathbf{U}\mathbf{U}^\intercal \right) \left(\mathbf{X} - \overline{\mathbf{x}}\right)\rVert_F,
\end{equation}
where $\overline{\mathbf{x}} \in\mathbb{R}^M$ is the empirical average of $\overline{\ell}$ over all the considered pressure-power combinations, the subscript $F$ denotes the Frobenius norm, and $\mathbf{I}$ is the identity.
The objective in Eq.~\eqref{eq:pod_obj} measures the error between the (mean-subtracted) data and its orthogonal projection onto the $r$-dimensional range of $\mathbf{U}$.
It can be shown that the optimal~$\mathbf{U}$ is given by the first $r$ left singular vectors of the data matrix $\mathbf{X} - \overline{\mathbf{x}}$, and the span of~$\mathbf{U}$ may be understood as the~$r$-dimensional subspace of maximum variance. 
Given~$\mathbf{U}$, the problem of learning $\overline{\ell}$ as a function of the parameters can be converted into an almost equivalent one, where the POD coefficients
\begin{equation}
\label{eq:POD_coeffs}
    \overline{c} \coloneqq \mathbf{U}^\intercal \overline{\ell} \in\mathbb{R}^r
\end{equation}
are learned as a function of $C$.
In order to fit the POD coefficients as a function of the parameters, we choose to use GPR, since it is well-known to perform well in the low-data limit, it is robust in the presence of noise, and it offers a measure of uncertainty via the posterior covariance. A brief mathematical description of GPR is given in App.~\ref{app:A1}.

\section{Results and discussion}\label{sec:results}
\subsection{Plasma jet behavior}
The time-averaged centerline profiles $\Bar{L}(x, P, W)$, obtained as discussed in Sec.~\ref{sec:jet_profile}, are here presented and discussed. 

\begin{figure}
    \centering
    \includegraphics[width=0.8\textwidth]{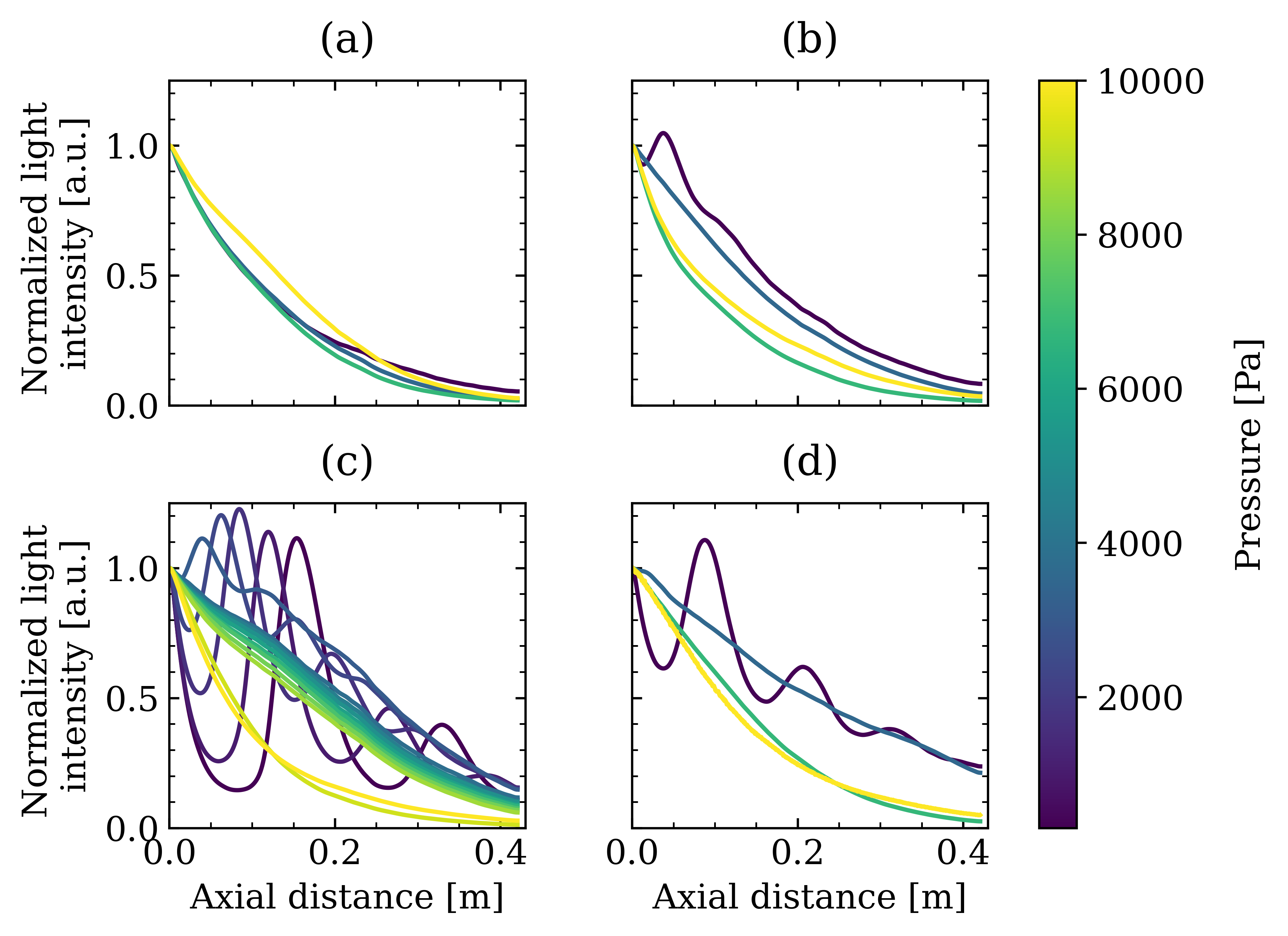}
    \caption{Emitted light $\Bar{L}(x, P, W)$ centerline profiles for a range of powers: (a) 100 kW; (b) 150 kW; (c) 200 kW; (d) 300 kW.}
    \label{fig:pressures}
\end{figure}

In Fig.~\ref{fig:pressures}, the jet profiles are shown as a function of chamber pressure for values of power of 100, 150, 200, and 300 kW.
At higher chamber pressures, the plasma jet consistently exhibits subsonic behavior across the entire range of torch powers. 
This can be inferred from the fact that there are no peaks (typically associated with shocks) in the jet profiles.
Conversely, Figs.~\ref{fig:pressures}(a)--(b) show that peaks in the profile arise when chamber pressures dip below 700 Pa, and torch powers exceed 140 kW. Fig.~\ref{fig:pressures}(c), illustrates the transition from supersonic to subsonic flow conditions resulting from changes in chamber pressure: as the reactor pressure increases, shock diamonds, which manifest as distinctive peaks of emitted light in the jet profiles, shift towards the torch exit. 
This shift correlates with a simultaneous increase in emitted light intensity on the first peak and a reduction in the number of shocks. 
Finally, Fig.~\ref{fig:pressures}(d) shows that even at high powers ($300$ kW, in this case), supersonic flow conditions can be achieved at low-enough chamber pressures (e.g., $600$ Pa).

\begin{figure}
    \centering
    \includegraphics[width=0.8\textwidth]{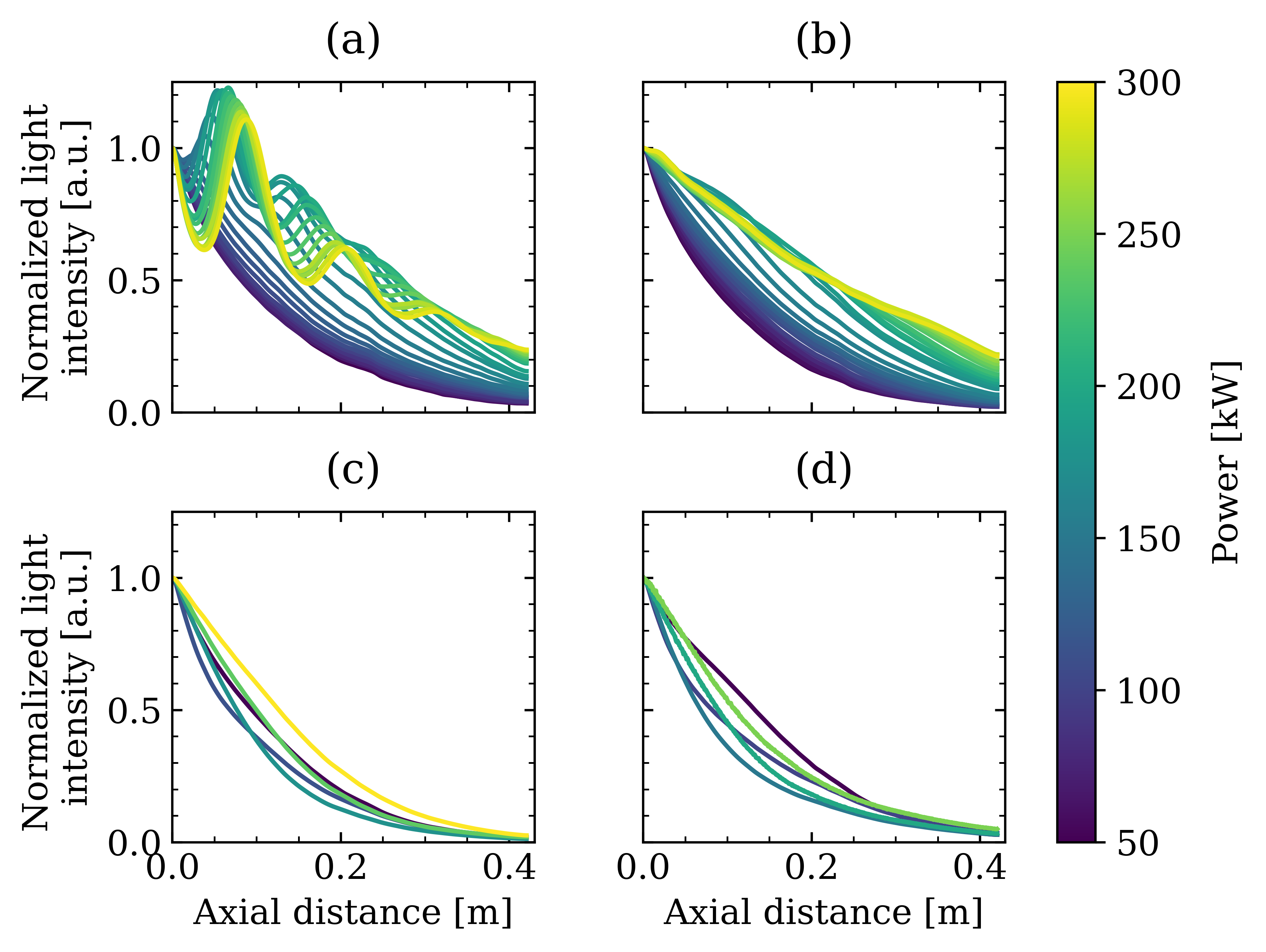}
    \caption{Emitted light $\Bar{L}(x, P, W)$ centerline profiles for a range of pressures: (a) 600 Pa; (b) 1000 Pa; (c) 5000 Pa; (d) 10000 Pa.}
    \label{fig:powers}
\end{figure}

Fig.~\ref{fig:powers} presents the jet profiles as a function of torch power for values of pressure of 600, 1000, 5000, and 10000 Pa.
Fig.~\ref{fig:powers}(a) demonstrates a transition from subsonic to supersonic flow conditions due to an increase in torch power. In this scenario, shock peaks migrate away from the torch exit towards the tail of the jet, accompanied by an overall amplification in emitted light intensity.
At relatively low pressures (Fig.~\ref{fig:powers}(b)), the emitted light from the plasma jet undergoes shape changes along its axial length. Specifically, higher power levels lead to an augmentation of emitted light at the tail of the jet, whereas lower power levels result in a more confined jet at the torch exit. This behavior is also encountered at higher chamber pressure (see Fig.~\ref{fig:powers}(c)-(d)).

\begin{figure}
    \centering
    \begin{subfigure}{0.45\textwidth}
        \includegraphics[width=1\textwidth]{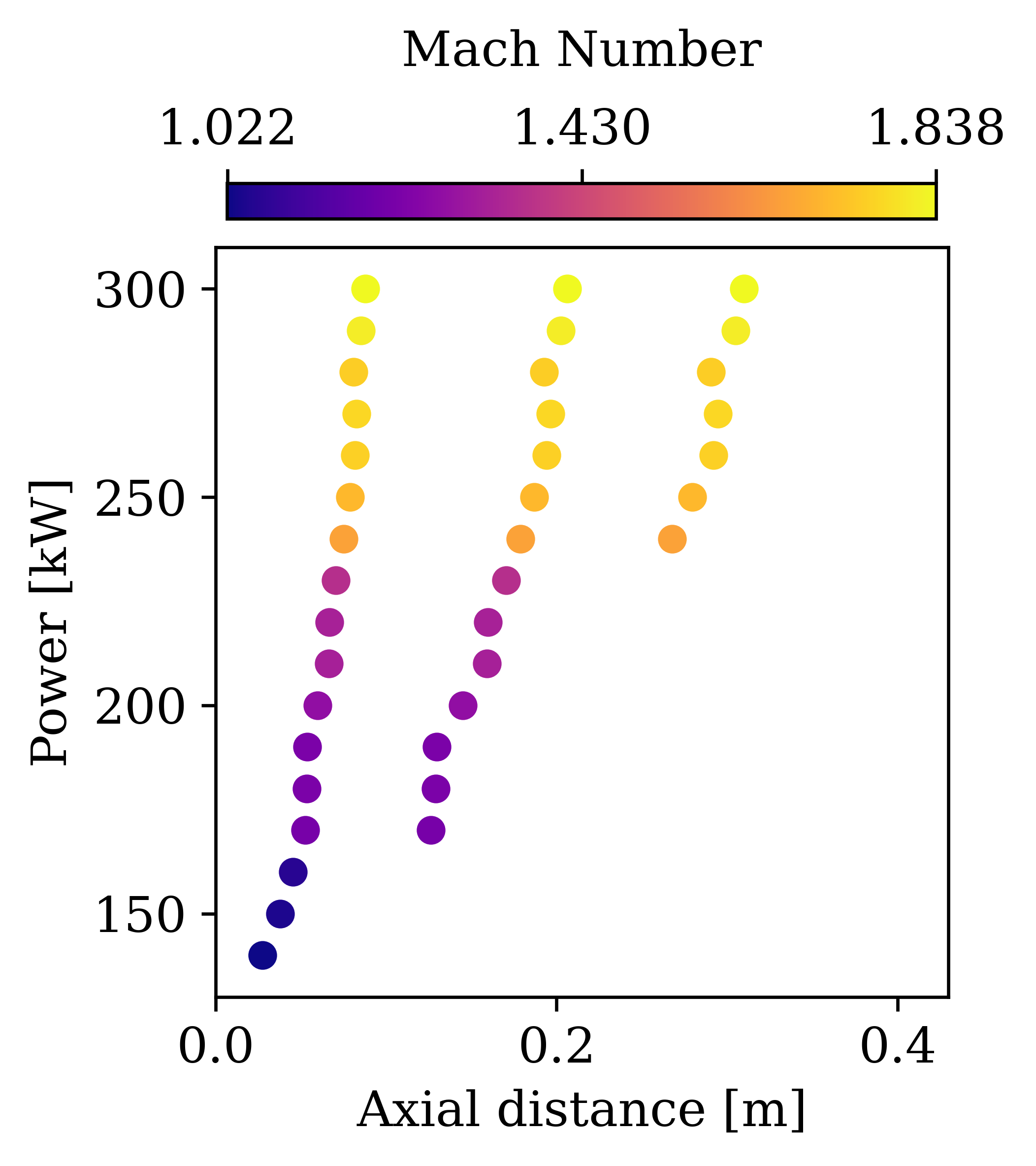}
        \subcaption{}
        \label{fig:--}
    \end{subfigure}
    \begin{subfigure}{0.45\textwidth}
        \includegraphics[width=1\textwidth]{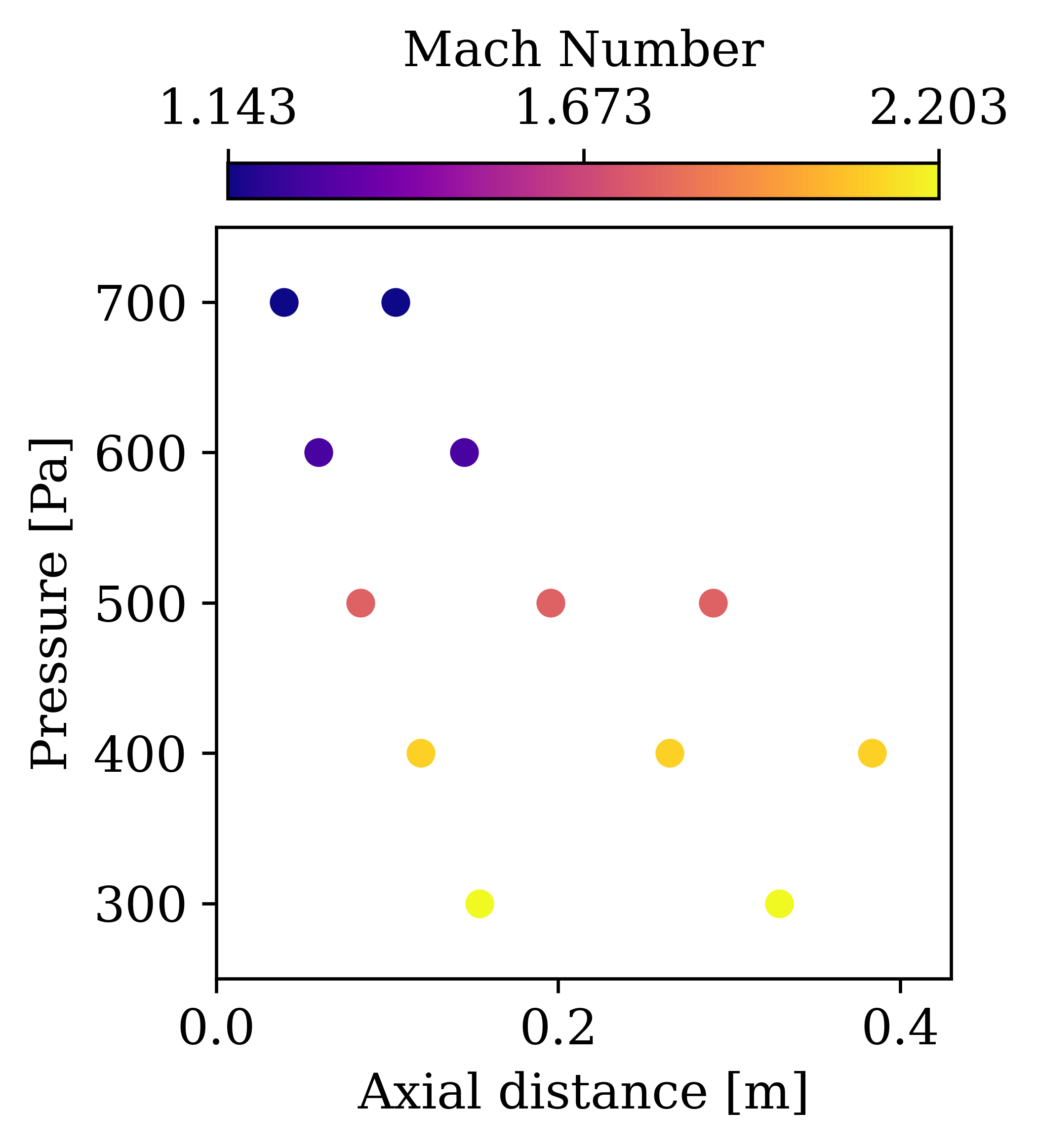}
        \subcaption{}
        \label{fig:--}
    \end{subfigure}
    \caption{Plasma jet shock locations: (a) Chamber pressure of 600 Pa, (b) ICP torch power of 200 kW. The Mach number is computed from Eq.~\eqref{eq:PP_eq}, with $D$ = 0.1 m (the diameter of the nozzle) and $l$ computed as the average distance between shocks.}
    \label{fig:peak_shift_mach}
\end{figure}

It is interesting to observe that supersonic flow conditions arise at given pressure/power combinations despite the use of a straight nozzle and a constant mass flow rate. 
This, however, should not be too surprising if we recall that, according to Rayleigh flow theory, heat addition (here obtained via a higher torch power and/or lower chamber pressure) drives a subsonic flow towards sonic conditions, and ultimately to a choked flow. 
Indeed, shocks are observed at high torch powers and low chamber pressure. 
Another noteworthy observation is that an increase in power and/or decrease in pressure lead to a corresponding increase in the number of shocks and in their average axial distance. 
This phenomenon can be correlated to a corresponding increase in the Mach number via the Prandtl-Pack theory~\citep{pack1950note}.
According to this theory, the axial distance between shocks in a supersonic jet is proportional to the jet Mach number $M_j$ through the relationship
\begin{equation}
    \frac{l}{D} = 1.306 \sqrt{M_j^2 -1},
\label{eq:PP_eq}
\end{equation}
where $l$ represents the average spacing between shock cells and $D$ is the diameter of the jet. The coefficient 1.306 comes from the ratio $\pi$/2.40483, the denominator being the first root of the Bessel function~\citep{powell2010prandtl}.
While we recognize that the Plasmatron X jet may not satisfy all the assumptions behind the Prandtl-Pack theory, we nonetheless use Eq.~\eqref{eq:PP_eq} to compute a rough estimate of the Mach number by taking $D = 0.1$ m (the nozzle exit diameter) and extracting $l$ from the data shown in Fig.~\ref{fig:peak_shift_mach}.
The computed Mach number is color-coded in Fig.~\ref{fig:peak_shift_mach}.

In order to obtain an alternative estimate of the Mach number, we also use the Rayleigh-pitot tube formula~\citep{anderson2000hypersonic}, which relates the Mach number to the pressure jump across a shock. 
The pressure jump is computed as the ratio of the post-shock pressure (obtained from the vibrational temperature computed using optical emission spectroscopy) and of the stagnation pressure, which we measure using a pitot probe. 
We find that the two approaches (Prandtl-Pack and Rayleigh-pitot) provide estimates that deviate from each other by about $12\%$ in the transonic regime and $6\%$ in supersonic flow conditions. 
We therefore consider the Mach number calculations displayed in Fig.~\ref{fig:peak_shift_mach} as a promising first estimate of the Mach number inside the Plasmatron X. 

\begin{figure}
    \centering
    \includegraphics[width=0.8\textwidth]{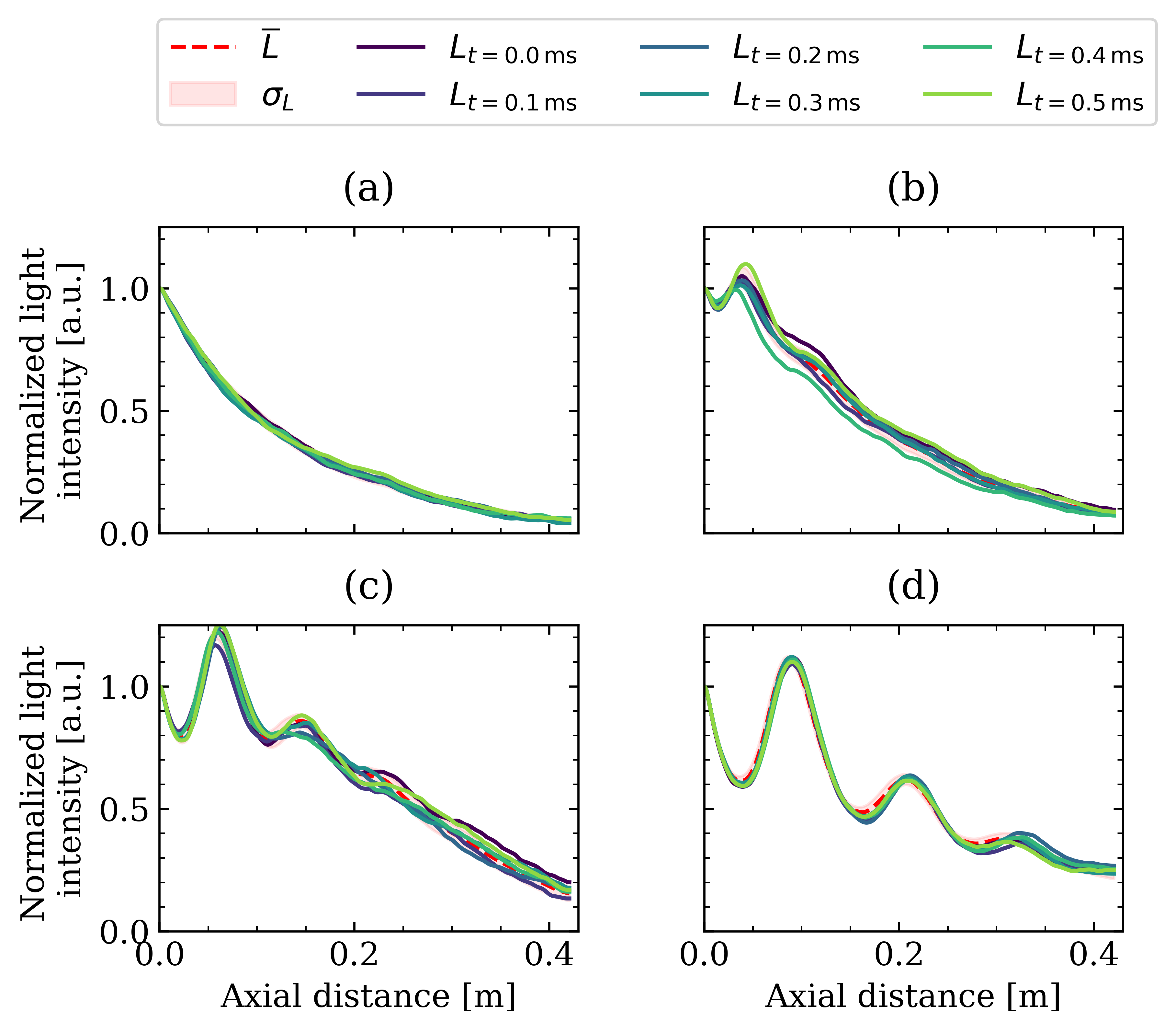}
    \caption{Temporal evolution of jet centerline profile with mean emitted light profile $\overline{L}$ and standard deviation $\sigma_L$ at 600 Pa and: (a) 100 kW, (b) 150 kW, (c) 200 kW, (d) 300 kW.}
    \label{fig:fast_prof1}
\end{figure}

\begin{figure}
    \centering
    \includegraphics[width=0.8\textwidth]{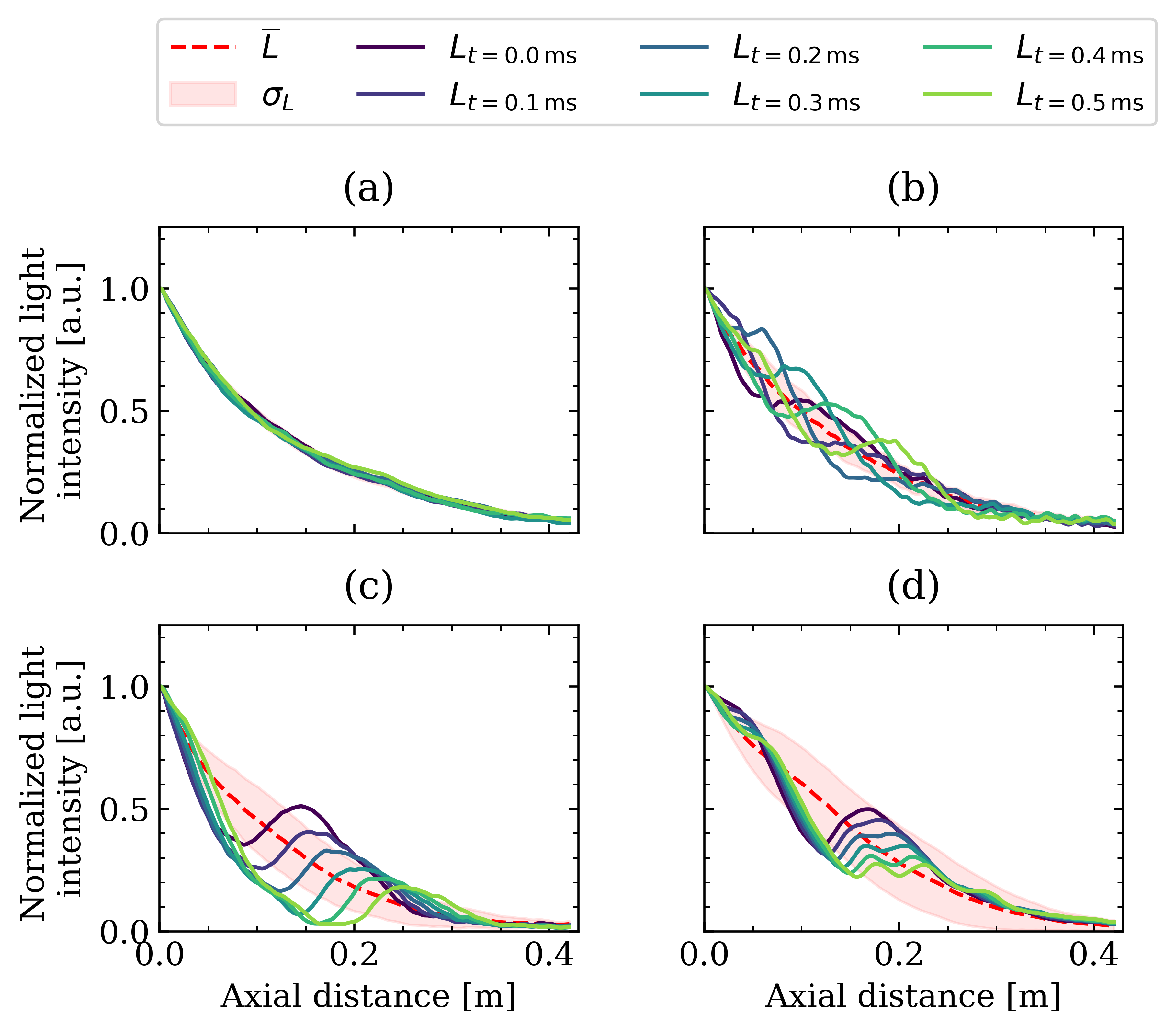}
    \caption{Temporal evolution of jet centerline profile with mean emitted light profile $\overline{L}$ and standard deviation $\sigma_L$ at 100 kW and: (a) 600 Pa, (b) 1000 Pa, (c) 5000 Pa, (d) 10000 Pa.}
    \label{fig:fast_prof2}
\end{figure}

Investigating potential instabilities in the Plasmatron X jet requires information on the temporal evolution of the jet profile. 
As a first approach, the temporal evolution at different torch powers and chamber pressures is visualized in Figs.~\ref{fig:fast_prof1}~--~\ref{fig:fast_prof2}, together with the mean profiles $\Bar{L}$ and the standard deviations $\sigma_L$ (68\% tolerance interval) evaluated over the entire data-acquisition window.
Fig.~\ref{fig:fast_prof1} reveals that, at low chamber pressure (i.e., 600 Pa), the profiles exhibit very low variance for all values of torch power (and especially at $100$ kW and $300$ kW). 
This implies that the jet is predominantly steady, with very low-amplitude perturbations about the mean profile.
In Fig.~\ref{fig:fast_prof2}, we keep power fixed at $100$ kW and we observe the effects of increasing the chamber pressure on the profile dynamics. 
In particular, we see that the standard deviation (pink strip) increases significantly as we increase the pressure, and this suggests that the temporal dynamics of the jet begin to exhibit high-amplitude oscillations about the mean.
Instantaneous profiles also reveal noteworthy features. 
At 1000 Pa, panel (b) suggests the presence of high-frequency structures and the evolution of the profiles is reminiscent of a \emph{standing} wave. 
At higher reactor pressure, we start to observe \emph{traveling}-wave behavior, with an increase in spatial wavenumber between panels (c) and (d), and a decrease in phase velocity. 

\begin{figure}
        \centering        
        \includegraphics[width=0.8\textwidth]{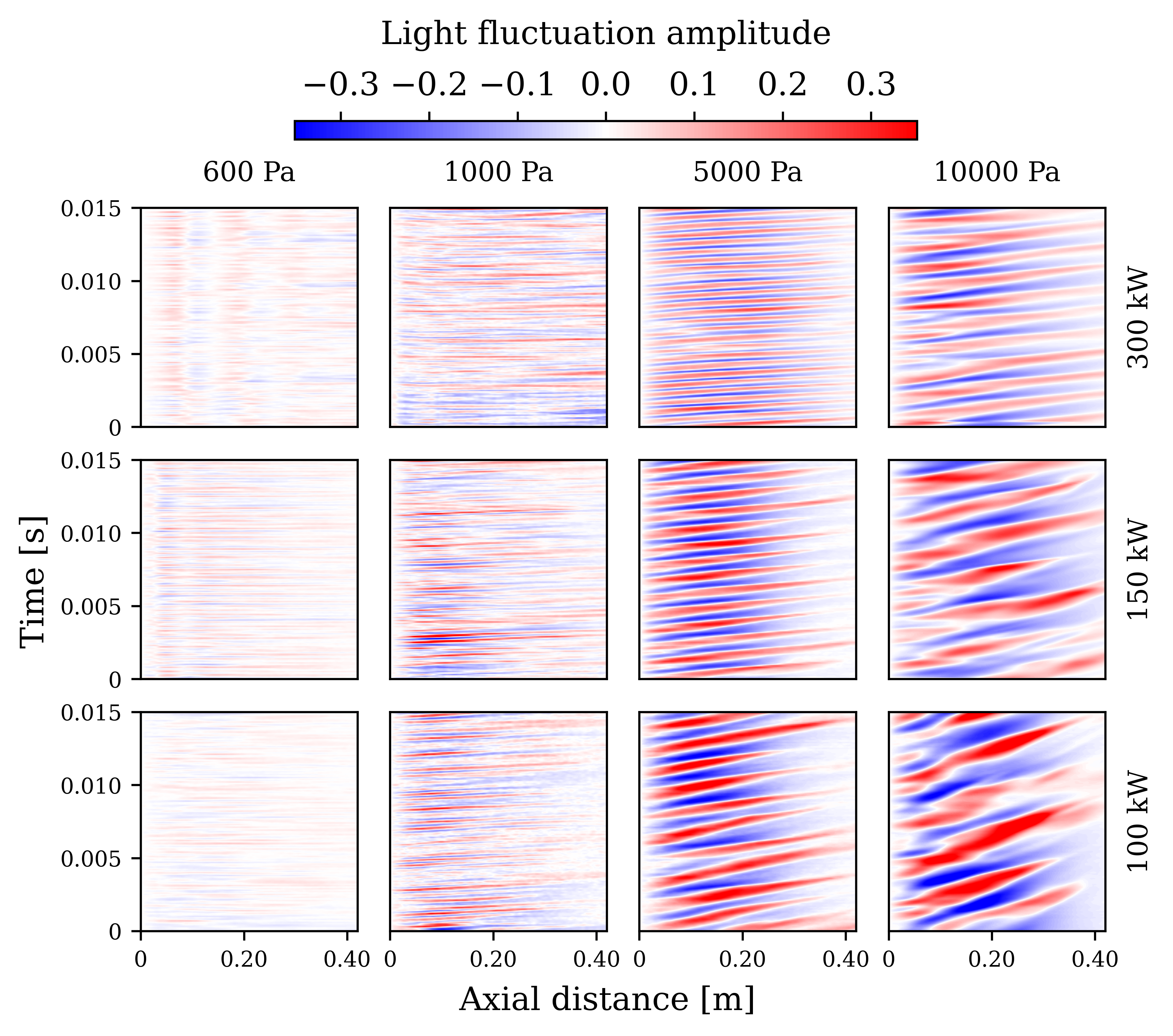}
        \caption{Emitted light fluctuation $L^\prime(x,t)$ in the space-time domain for selected combinations of pressure $P$ and power $W$.}
        \label{fig:time_sub}
\end{figure}

\begin{figure}
    \centering
    \begin{subfigure}{0.49\textwidth}
        \includegraphics[width=1\textwidth]{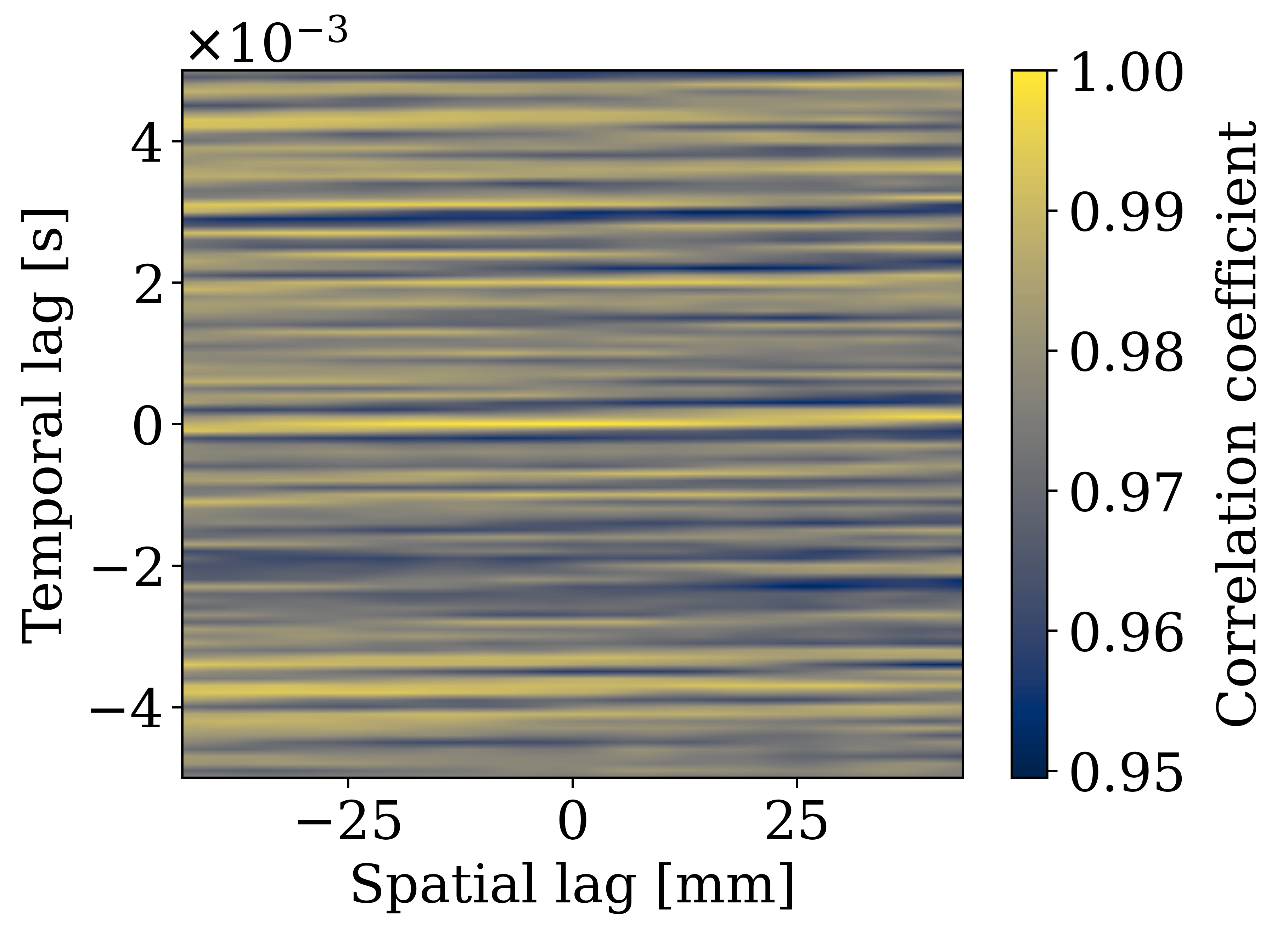}
        \subcaption{}
        \label{fig:pod_600}
    \end{subfigure}
    \begin{subfigure}{0.49\textwidth}
        \includegraphics[width=1\textwidth]{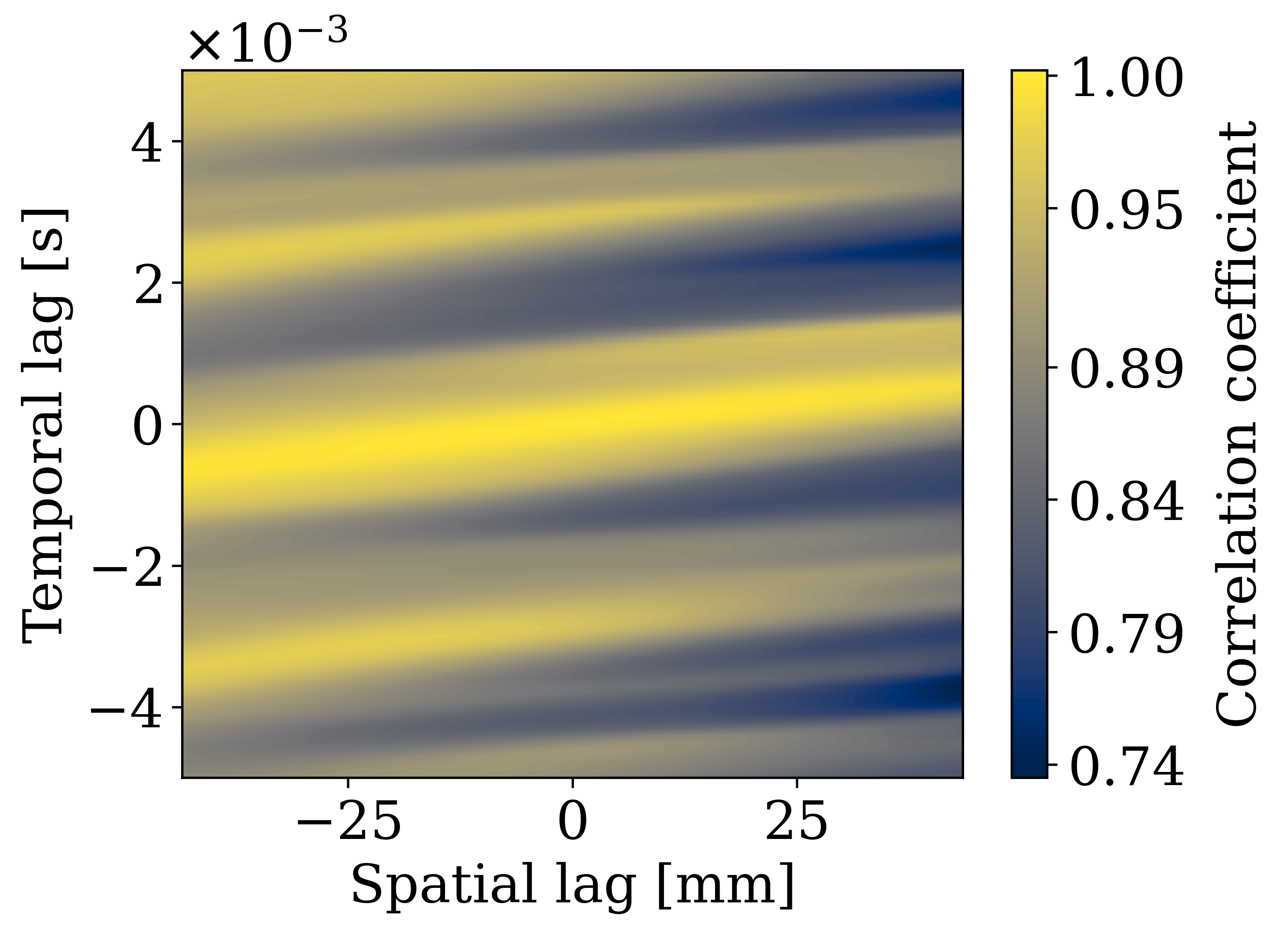}
        \subcaption{}
        \label{fig:pod_1000}
    \end{subfigure}
    \caption{Space-time correlation analysis performed at $x_0$~=~0.15~m and $t_0$~=~0.01~s, and at 100 kW torch power: (a) 100 Pa and (b) 10000 Pa. The difference in the colorbar limits is necessary to appreciate $R_{LL}$ distributions.}
    \label{fig:STcorr}
\end{figure}

This analysis is extended to a larger set of $P$-$W$ combinations of interest, through the space-time approach discussed in Sec.~\ref{sec:jet_profile}, and the results are shown in Fig.~\ref{fig:time_sub}.
Across the full range of power settings, negligible fluctuations in amplitude persist throughout the temporal evolution of the jets at the lowest pressure, as shown by the left-hand panels of Fig.~\ref{fig:time_sub}. During this phase, no discernible dynamic effect can be observed. 
As the chamber pressure rises (e.g., at 1000 Pa), we observe larger-amplitude (and higher-frequency) perturbation about the mean, particularly in the proximity of the torch. 
At 5000 Pa, the amplitude escalation persists, predominantly at lower power levels. Additionally, the fluctuations appear to lower in frequency, leading to broader blue-red regions. Peaks and troughs of equal amplitude span various axial positions at distinct temporal moments, resulting in the transformation of iso-amplitude regions from horizontal to oblique configurations.
While this phenomenon is generally well-defined at 5000 Pa, when the pressure is increased to 10000 Pa (as illustrated on the right-hand side of Fig.~\ref{fig:time_sub}), a similar pattern emerges but with a comparatively higher degree of complexity. 
Amplitudes are largest, and the oblique areas expand further. Notably, spatial and temporal disturbances become evident both at the torch exit and the tail of the plasma jet.
The effect of the pressure increase is quantified in terms of fluctuation-regions slope change in Fig.~\ref{fig:time_sub}. The slope is in fact proportional to the to phase speed of the emitted light fluctuation waves. The most interesting trend is observed at 100 kW, where the average fluctuation slope evolves from about 0.102 ms/m at 600 Pa, to 0.496 ms/m at 1000 Pa, 5.516 ms/m at 5000 Pa, reaching its maximum value of 11.874 ms/m at the highest pressure of 10000 Pa.
The same behavior can also be appreciated in Fig.~\ref{fig:STcorr}, which shows the correlation (Eq.~\eqref{eq:stcorr}) computed at $x_0$~=~0.15~m and $t_0$~=~0.01~s (i.e., mid-stream axial location) for two different pressures at the same torch power. 
The values of $x_0$ and $t_0$ were chosen as 10\% and 5\% of the data domain, respectively.
Computing the correlation for other values of $x_0$ and $t_0$ (not shown) did not change the qualitative nature of the results.

\begin{figure}
        \centering        
        \includegraphics[width=0.8\textwidth]{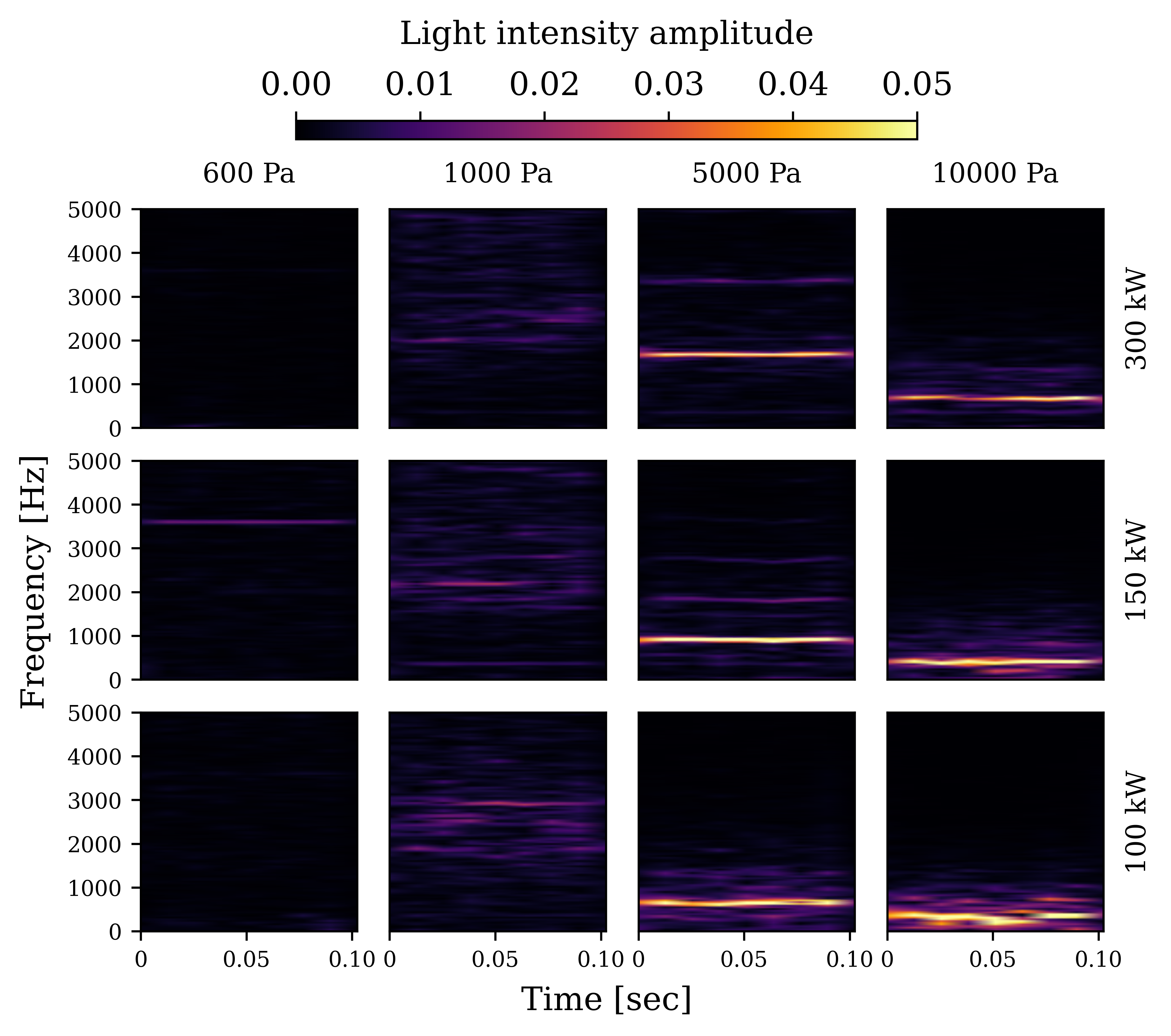}
        \caption{Short-Time-Fourier-Transform of emitted light fluctuations $L^\prime(x,t)$ at $x$ = 0.15 m axial location.}
        \label{fig:stft}
\end{figure}

For completeness, we also study the frequency-domain signature of the dynamics of the jet profiles.
In particular, Fig.~\ref{fig:stft} shows the STFT of emitted light fluctuations $L^\prime(x,t)$ at $x$~=~0.15~m axial location for same $P$-$W$ combinations. 
This figure confirms the behavior described so far, with low-amplitude fluctuations at low pressures and large-amplitude fluctuations at higher pressures. 
We also point out an interesting feature: at high pressures (5000 and 10000 Pa), the frequency-content of the fluctuations changes significantly as a function of power. 
At higher powers, the fluctuations are predominantly monochromatic, while the number of active frequencies increases at lower pressures. 
Moreover, we observe that as we increase the power, the dominant frequency of the fluctuations also increases (leading to faster oscillations). 
(Analogous results are obtained at different axial locations.)

The appearance of traveling waves along the free stream plasma jet profile can be attributed to large-scale convective effects, which result from fluid instabilities formed between the plasma jet and the surrounding gas as the chamber pressure is changed~\citep{clemens1995large, danaila1998mode}. More specifically, the density and velocity distributions of the surrounding gas and plasma are changed as the chamber pressure is increased. When a significant difference between their velocities arises, a shear layer is formed at the interface between the plasma and the background gas~\citep{michalke1984survey}. The shear layer experiences disturbances and becomes unstable, leading to the formation of convective effects that we here observe as traveling waves along the jet profiles. The instability can be further enhanced by factors such as density gradients, temperature differences, and surface irregularities~\citep{papamoschou1988compressible, corke1991mode}. These factors contribute to the development and propagation of the waves, which can manifest as sinuous or roll-up vortical structures along the jet~\citep{danaila1998mode, cohen1987evolution}.

\subsection{Prediction of emitted light profiles for unseen operating conditions}
\begin{figure}
    \centering
    \begin{subfigure}{0.49\textwidth}
        \includegraphics[width=1\textwidth]{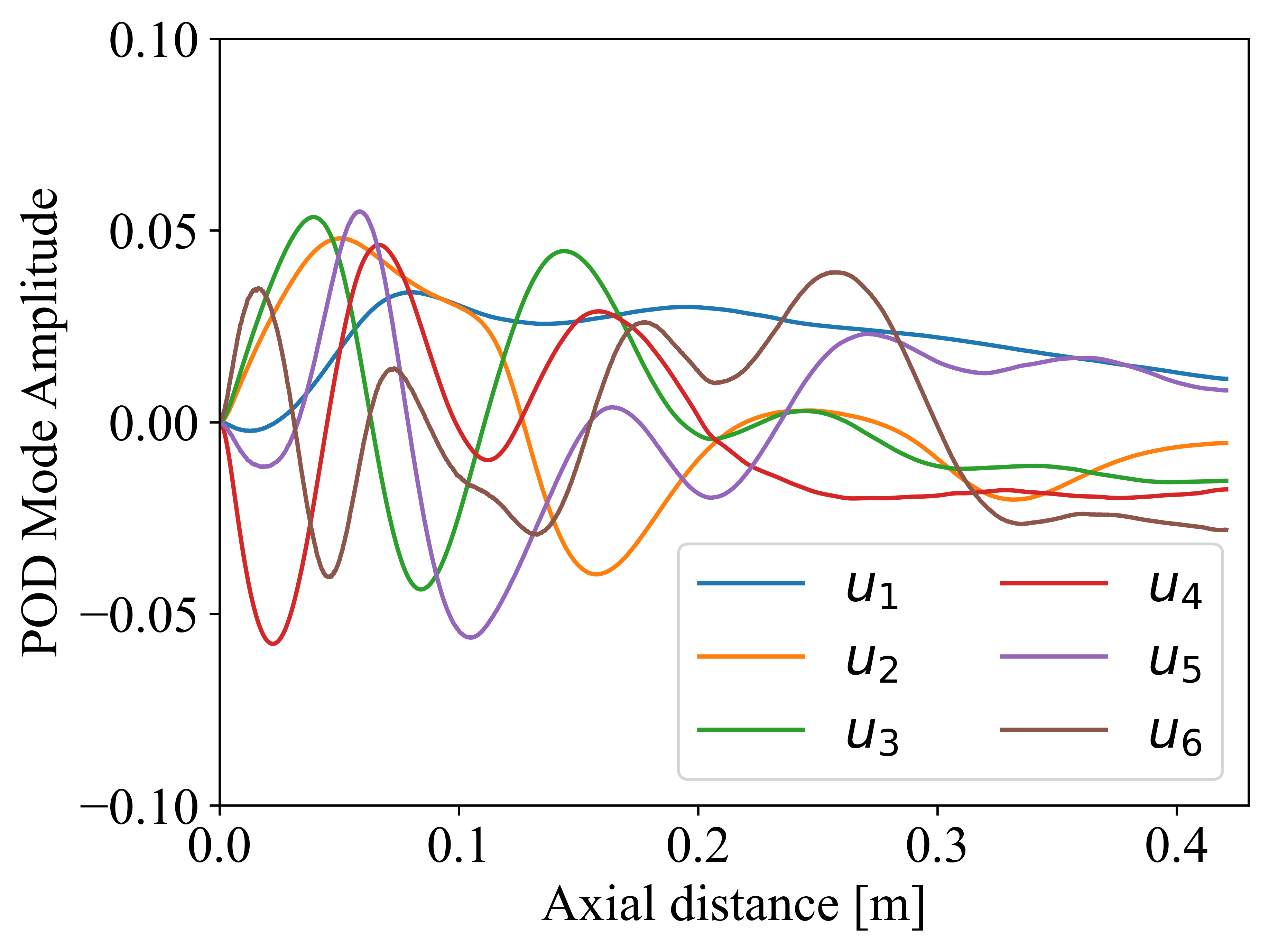}
        \subcaption{}
        \label{fig:modes}
    \end{subfigure}
    \begin{subfigure}{0.49\textwidth}
        \includegraphics[width=1\textwidth]{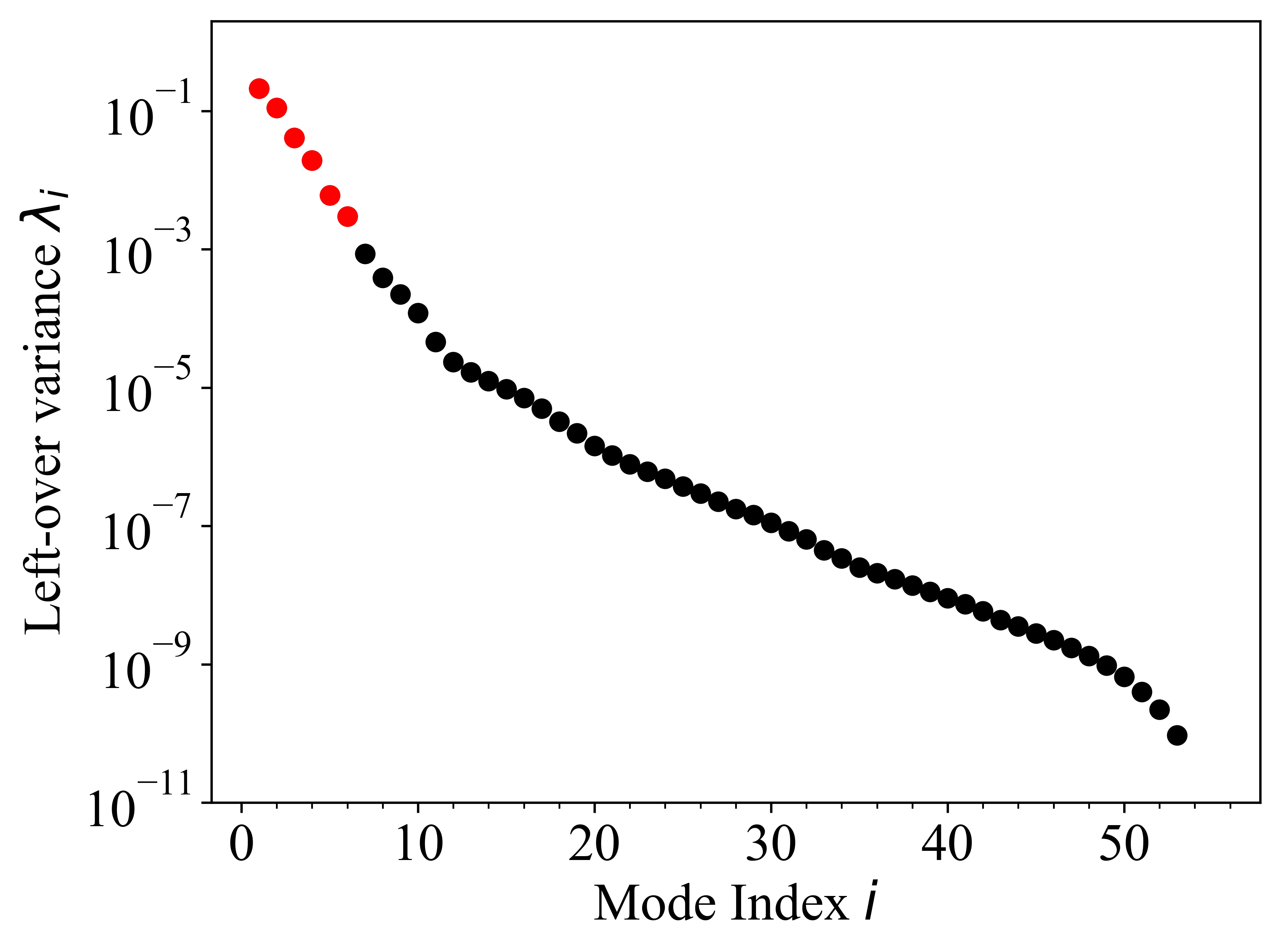}
        \subcaption{}
        \label{fig:lambda}
    \end{subfigure}
    \caption{Results of SVD: (a) POD modes $u_i$ and (b) left-over variance $\lambda_i$ (red markers are the first modes $u_{1-6}$).}
    \label{fig:pod}
\end{figure}

\begin{figure}
    \centering
    \begin{subfigure}{0.49\textwidth}
        \includegraphics[width=1\textwidth]{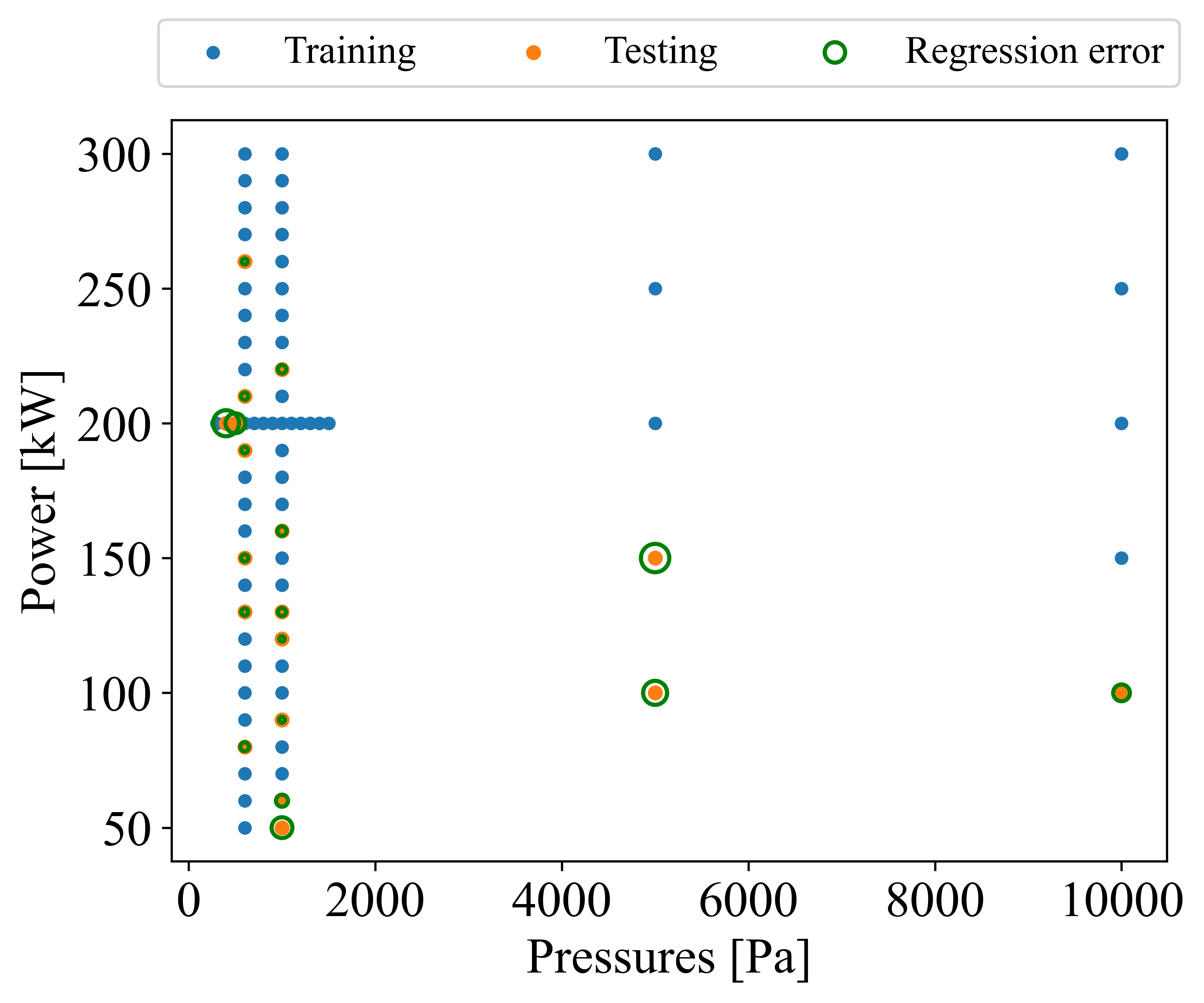}
        \subcaption{}
        \label{fig:domain}
    \end{subfigure}
    \begin{subfigure}{0.49\textwidth}
        \includegraphics[width=1\textwidth]{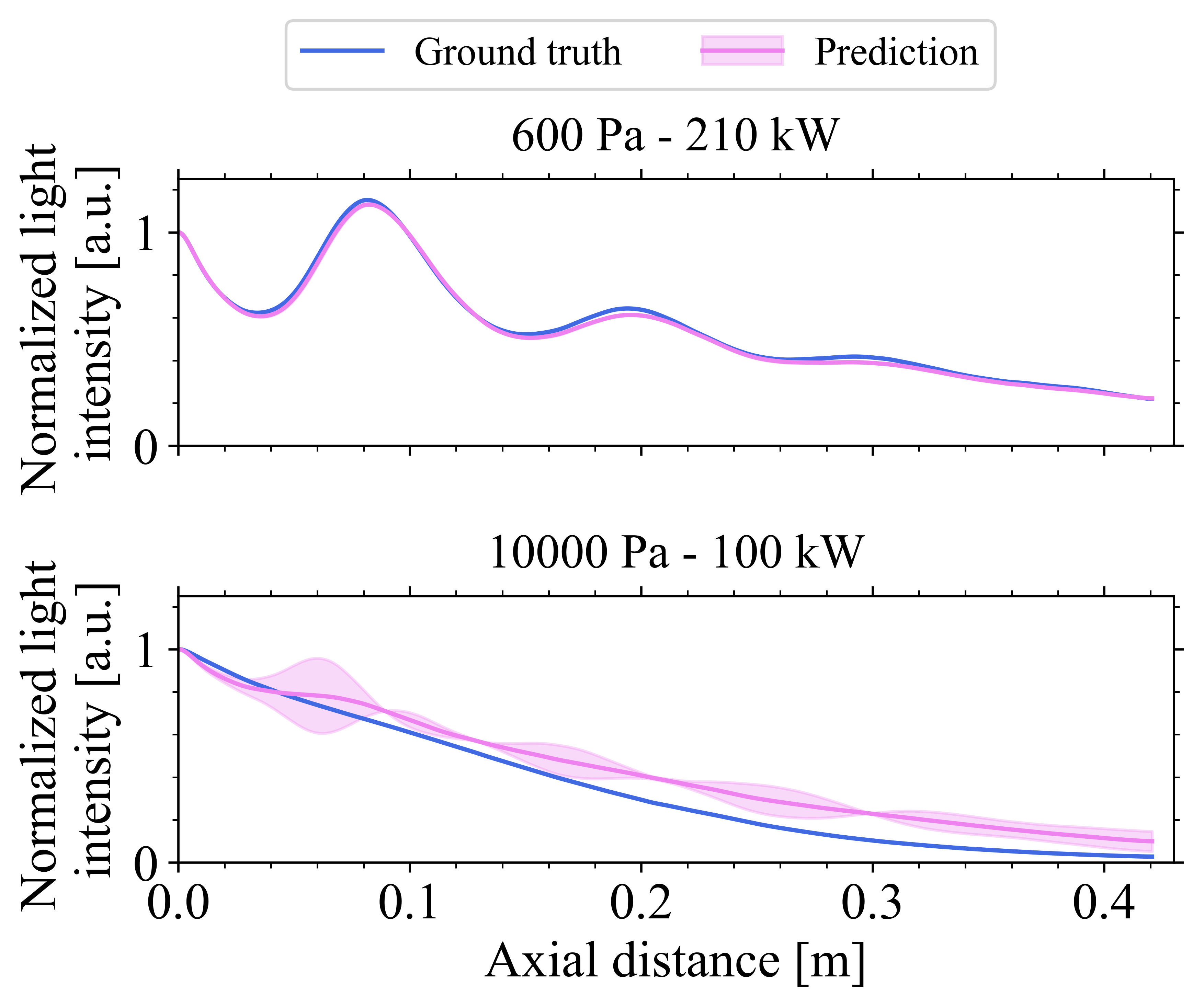}
        \subcaption{}
        \label{fig:gpr2d2}
    \end{subfigure}
    \caption{Results of GPR prediction: (a) domain of training and testing points with regression uncertainty; (b) ground truth and prediction results of jet profiles in physical space.}
    \label{fig:gpr_results}
\end{figure}

In Sec.~\ref{sec:gpr_method}, we discussed our approach for learning jet profiles $\ell$ as a function of pressure and power. 
Our experimental data set includes jet profiles at 73 pressure/power combinations. We randomly allocated 55 (75\%) for training and reserved 18 for testing.
We found that the first 6 POD modes (Fig.~\ref{fig:pod}(a)) capture 99.9\% of the variance. 
A plot of the left-over variance 
\begin{equation}
\label{eq:lambda_i}
    \lambda_i = 1 - \frac{\sum_{j=1}^{i}\sigma_j^2}{\sum_{j=1}^N\sigma_j^2}
\end{equation}
is shown in Fig.~\ref{fig:pod}(b).
Here, $\sigma_j$ is the $j$th singular value and $N$ is the total number of snapshots in the data matrix (see Eq. \eqref{eq:pod_obj}).
Since the leading 6 POD modes capture most of the variance of the data set, we used GPR to fit the first 6 POD coefficients as a function of pressure and power.
The obtained model is showcased in Fig.~\ref{fig:gpr_results}. 
Fig.~\ref{fig:gpr_results}(a) shows the training (blue) and testing (orange) points, while the green circles illustrate the prediction error associated with the first coefficient $\overline{c}_1$. 
The diameter of the circles scales proportionally to the relative error
\begin{equation}
    e = \frac{\lVert\overline{L}_{true} - \overline{L}_{GPR}\rVert_2}{\lVert\overline{L}_{true}\rVert_2},
\end{equation}
where $\lVert \cdot\rVert_2$ denotes the two norm. 
As expected, the error is higher (i.e., bigger diameter) for testing points "far" from training points, and vice-versa. 
For reference, the minimum and maximum values found for $e$ are 0.0044 and 0.0126, respectively, corresponding to percent errors of $0.44$ and $1.26$.

The learned POD coefficients can then be used to reconstruct time-averaged jet profiles in physical space, presented for two of the testing conditions in Fig.~\ref{fig:gpr_results}(b).
At $P = 600$ Pa and $W = 210$ kW (top panel), the prediction is quite accurate and the regression uncertainty (measured by the thickness of the pink region) is low.
This is most certainly due to the fact that the testing point $(P,W) = (600,210)$ lies in a region that was heavily sampled during training (see Fig.~\ref{fig:gpr_results}(a)), and we therefore expect a good performance. 
Conversely, at $P = 10^{4}$ Pa and $W = 100$ kW (bottom panel), the predicted profile is less accurate (although not terrible) and the regression uncertainty is higher. 
This is probably due to the fact that the testing point $(P,W) = (10^{4},100)$ lies outside the training region. 

\section{Conclusions}\label{sec:conclusions}
Characterizing the test environment of plasma wind tunnels is the key for properly replicating relevant entry conditions experienced by TPS materials during hypersonics flights. This research focused on the Plasmatron X ICP torch wind tunnel, investigating the influence of the input torch power and the reactor chamber pressure on the behavior of its plasma jet. This has been achieved by acquiring high-speed images of the visible-spectrum-light emitted by the jet under different system configurations, and by studying the centerline axial profile features with space, time, and frequency domain approaches. While in supersonic jet conditions (relatively low chamber pressure), the torch power determines the location, the width and the intensity of shock diamonds, in subsonic conditions the reactor chamber pressure dominates the dynamical effects of jet. In fact, as the pressure increases, the flow characteristics change and fluid instabilities arise, leading to large-scale convective effects that can be observed and quantified in the space-time and frequency domains. Moreover, given the relevance of plasma jet profile, and the data-acquisition cost/time required to cover the extensive operational envelope of the Plasmatron X wind tunnel, this research demonstrated that the synergy between a machine learning paradigm and a data-driven decomposition method was capable of accurately predicting jet profiles at unseen pressure-power combinations.

The findings of this research contribute to the understanding the underlying physics of ICP plasma jets, giving a solid support for the design phase of material testing and diagnostics campaigns, as well as helping in the validation of accurate simulation models. Future works will extend the analysis to newly collected conditions, investigating the influence on the jet dynamics of other parameters (nozzle geometry and mass flow rate), in nitrogen and carbon dioxide plasma jets.

\backmatter
\bmhead{Acknowledgments}
This research was supported by The Grainger College of Engineering at the University of Illinois Urbana-Champaign.
L.C. is supported by NASA under award number 80NSSC21K1117.
A.P. is supported by the National Science Foundation under grant number 2139536.
The authors acknowledge Matthew Konnik and Trey Oldham for having supported the experimental campaign of this research.
\bmhead{Declarations}
\begin{itemize}
\item Conflict of interest: All authors declare that they have no conflicts of interest.
\item Availability of data and materials: The data are available from the corresponding author, upon reasonable request.
\item Authors' contributions: L.C.: conceptualization, data curation, formal analysis, investigation, methodology, software, validation, visualization, supervision, writing original draft, writing review and editing; 
A.P.: conceptualization, formal analysis, methodology, software, validation, visualization, supervision, writing original draft, writing review and editing; 
G.E, M.P., D.B., and F.P.: project administration, resources, supervision, writing review. 
\end{itemize}

\begin{appendices}

\section{Gaussian Process Regression}\label{app:A1}
In general, a Gaussian process can be understood as a collection of jointly-Gaussian random variables~\citep{williams2006gaussian}. 
In the context of Gaussian process regression, it is assumed that the data $\{\overline{c}_i\}_{i=1}^N$ (in our specific case, the POD coefficients in Eq.~\eqref{eq:POD_coeffs}) are generated by a function
\begin{equation}
    f: C \mapsto \overline{c},
\end{equation}
drawn from a Gaussian distribution with mean $\mu(C) = \mathbb{E}\left[f(C)\right]$ and covariance $K(C,C^\prime) = \mathbb{E}\left[\left(f(C)-\mu(C)\right)\left(f(C^\prime)-\mu(C^\prime)\right)\right]$.
That is,
\begin{equation}
\label{eq:gpr_prior}
    f(C) \sim \mathcal{N}\left(\mu(C),K(C,C^\prime)\right).
\end{equation} 
Regression is performed by leveraging the well-known result that the marginal distribution of jointly-Gaussian random variables is also Gaussian~\citep{williams2006gaussian}, so that we may write
\begin{equation}
\label{eq:gpr_posterior}
    f(C)\lvert f(\widetilde{C}) \sim \mathcal{N}\left(\mu(C)\lvert f(\widetilde{C}) ,K(C,C^\prime)\lvert f(\widetilde{C}) \right).
\end{equation}
In words, given normally-distributed observations $f(\widetilde{C})$, the function $f(C)$ is drawn from the conditional (posterior) distribution (Eq.~\eqref{eq:gpr_posterior}) with posterior mean $\mu(C)\lvert f(\widetilde{C})$ and covariance $K(C,C^\prime)\lvert f(\widetilde{C})$. 
The functional form of the posterior mean and covariance can be found, e.g., in Eq.~(2.19) in \cite{williams2006gaussian}.
In practical applications, the prior mean $\mu(C)$ in Eq.~\eqref{eq:gpr_prior} is typically taken to be~$0$, and the covariance $K(C,C^\prime)$ is defined via a user-defined kernel.
The choice of kernel should reflect the a-priori knowledge that we have of the underlying process that generated the data (i.e., the plasma jet in our case). 
For instance, it is reasonable to assume that the function $f(C)$ that we seek is continuous and differentiable with respect to the input~$C$. 
Throughout, we specify the covariance via the radial basis function kernel as follows,
\begin{equation}
\label{eq:kernel}
    K(C,C^\prime) = \exp{-\frac{1}{2\sigma^2}\lVert C - C^\prime\rVert^2},
\end{equation}
where the length-scale parameter $\sigma$ is to be optimized. 
Interestingly, Eq.~\eqref{eq:kernel} shows that the covariance of the output $f(C)$ is written as a function of the input~$C$. 
Intuitively, this implies that ``nearby" inputs $C$ produce ``nearby" outputs $f(C)$, while ``far-away" inputs lead to ``far-away" outputs. 
This can be loosely understood as a statement of continuity, so that it becomes intuitively clear how the chosen kernel defines a distribution from which we draw continuous, and infinitely-many-times differentiable, functions $f$. 
\end{appendices}

\bibliography{biblio}
\end{document}